\documentclass[preprint,12pt]{elsarticle}

\usepackage{amssymb}
\usepackage{amsmath}
\usepackage{url}

\journal{Measurement}

\begin{document}

\begin{frontmatter}



\title{Measurement of the intrinsic sensitivity for a single-ended accelerometer without the influence of the mounting condition}


\author{Tomofumi Shimoda\corref{cor1}}
\author{Wataru Kokuyama}
\author{Hideaki Nozato}
\cortext[cor1]{Corresponding author (e-mail: tomofumi.shimoda@aist.go.jp)}
\affiliation[label1]{organization={National Metrology Institute of Japan, National Institute of Advanced Industrial Science and Technology},%Department and Organization
            addressline={1-1-1 Umezono}, 
            city={Tsukuba},
            postcode={305-8563}, 
            state={Ibaraki},
            country={Japan}}

\begin{abstract}
The calibration technique for accelerometers has been internationally developed for up to 20~kHz to ensure the reliability of vibration measurement. 
However, it has been established that the calibrated sensitivity changes at over 10~kHz depending on the mounting conditions, and this makes it difficult to accurately measure the characteristics of accelerometers and degrades the accuracy of high-frequency vibration measurements. 
Thus, in this study, we developed a reversed-calibration method for measuring the intrinsic sensitivity of an accelerometer without the influence of the mounting conditions. 
Through demonstration experiment, the intrinsic resonance structure of the accelerometer at approximately 45.8~kHz was adequately determined. 
Furthermore, the result was independently confirmed by fitting the conventional adapter-calibration results up to 100~kHz with four different materials based on the dynamic three-body model. 
Concurrently, the material dependency observed during adapter calibration was quantitatively analyzed, after which its relationship with the Young's modulus was extracted. 
Overall, these results deepen our understanding of the performance of the accelerometer at above 10~kHz, which is essential in vibration metrology and accelerometer development.
\end{abstract}



\end{frontmatter}


\section{Introduction} \label{sec:introduction}
Accelerometer calibration is a key technological element for ensuring the reliability of vibration measurements in various industrial fields. 
National metrology institutes (NMIs) typically perform vibration calibration through a chain of primary calibrations and comparison calibrations using calibrated accelerometers. 
Generally, primary calibration is performed, following ISO16063-11 \cite{ISO16063-11}.
In this method, the accelerometer-output signal is compared with the reference vibration measured by a laser interferometer. 
In recent responses to the increasing demand for high-frequency vibration measurements, NMIs globally have developed primary calibration technologies that handle up to 20 kHz \cite{Dobosz1997I,Dobosz1997II,Martens2000,Martens2013,Winther2022,Kokuyama2022}.
International comparisons, such as CCAUV.V-K2 (10~Hz--10~kHz) \cite{CCAUV.V-K2} from 2009 to 2011 and CCAUV.V-K5 (10~Hz--20~kHz) \cite{CCAUV.V-K5} from 2017 to 2019, have also been conducted to ensure international equivalence, achieving equivalence within a few percent up to 20~kHz.

In the existing primary calibration for single-ended (SE) accelerometers, the accelerometers are mounted on a stainless-steel adapter, following Ref. \cite{Ripper2013}.
Thereafter, the laser beam of the reference interferometer is pointed to the upper surface of the adapter (as shown in Fig.~\ref{fig:adaptercal_schematic} later), as the sensing surface of the accelerometer is attached to the adapter, rendering it inaccessible to the laser.
However, the calibrated sensitivity obtained using this setup depends on the mounting conditions, including the target material and surface roughness \cite{Taubner2010,Bruns2012}. 
This is because of the differences in the vibrations at the sensing surface of the accelerometer and the surface of the adapter surface attributed to elastic deformation and contact stiffness \cite{Kokuyama2022,Bruns2012}. 
Thus, the existing calibration method can only measure the sensitivity under a specific mounting condition. 
Therefore, it is desirable to develop a method that does not depend on the mounting conditions.

Here, we developed a reversed-calibration method for measuring the sensitivity of an accelerometer without relying on the influence of the mounting conditions. 
In this method, the interferometer laser is directly pointed to the sensing surface of the accelerometer using a reversed-calibration jig, thus circumventing the effect of the deformation of the mounting surface on the reference vibration and ensuring that the measured sensitivity is intrinsic to the accelerometer. 
Notably, such intrinsic sensitivity is essential for determining the performance of the accelerometer performance without the effect of the measurement setup. 
Additionally, the influence of the mounting condition in the existing calibration method can be quantitatively evaluated based on the intrinsic sensitivity. 
We also performed the conventional adapter calibration to validate the reversed-calibration result via analyses based on a dynamic three-body model. 
Furthermore, the detailed dependency of the measured sensitivity on the mounting conditions was observed up to 100~kHz. 
Notably, adapter calibration above 50 kHz was performed for the first time in this study.

The remainder of this paper is organized, as follows: 
Section~\ref{sec:model} introduces the dynamic three-body model employed for all the analyses;
Section~\ref{sec:reversed} describes the reversed-calibration method for measuring intrinsic sensitivity;
Section~\ref{sec:adapter} discusses the adapter-calibration method;
Section~\ref{sec:result} presents the calibration results;
and Section~\ref{sec:discussion} discusses the results and concludes the study.

\section{Three-body model of an accelerometer} \label{sec:model}
\begin{figure}
	\begin{center}
	\includegraphics[width=10cm]{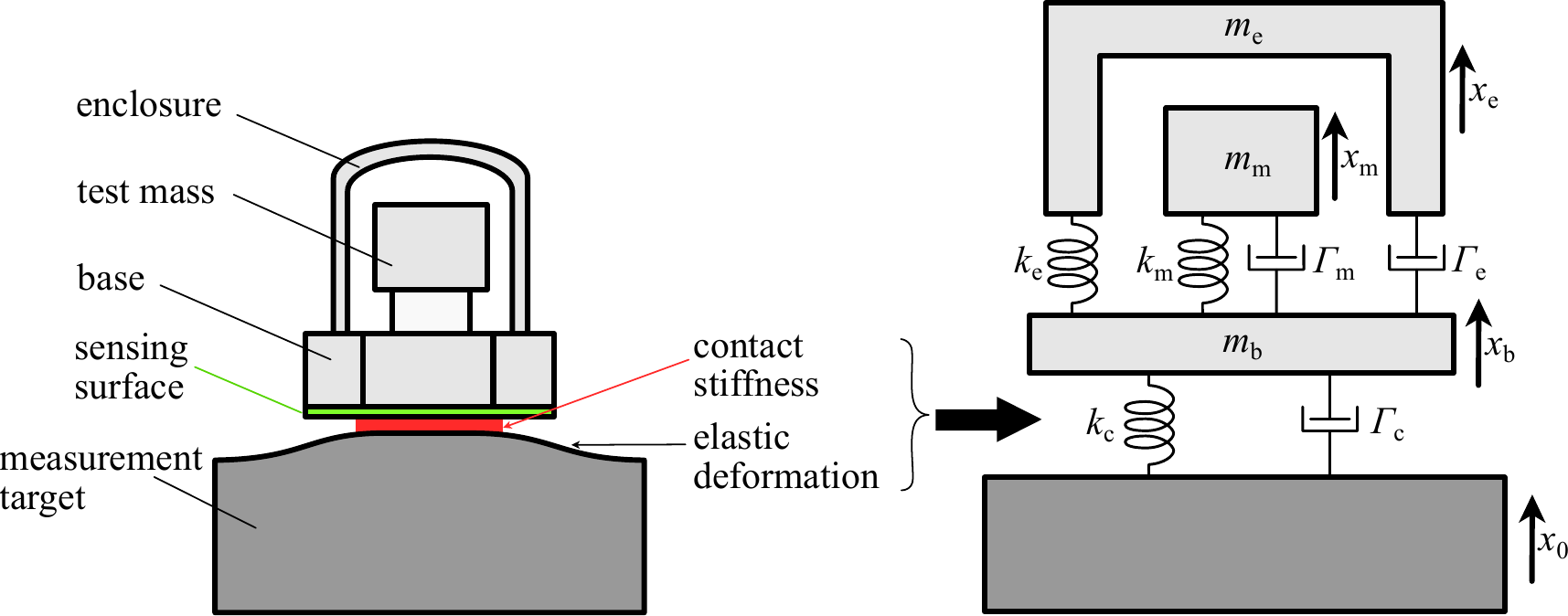}
	\caption{Mechanical model of an accelerometer. }
	\label{fig:model}
	\end{center}
\end{figure}
Here, we introduce a dynamic model for the accelerometer to elucidate the mounting-condition-dependent sensitivity changes. 
The mechanical part of a typical standard accelerometer comprises a seismic system and an enclosure. In the extant studies (e.g. \cite{Taubner2010,Bruns2012}), an accelerometer was modeled as a series mass–spring system comprising the base and seismic mass; thus, it can be referred to as a two-body model. 
In this study, we considered the effect of the enclosure (Fig.~\ref{fig:model}) to analyze the sensitivity of the accelerometer up to 100~kHz. 
The modified model is referred to as a three-body model throughout this paper. 
The elastic deformation of each component is simplified as the elongation of the spring, enabling the analytical calculation of the sensitivity. 
This simplification implies that only the lowest-order deformation mode of each part is considered, making it valid up to approximately 100~kHz.

The equations of motion of the masses are described, as follows:
\begin{eqnarray}
m_\mathrm{m} \ddot{x}_\mathrm{m} &=& -k_\mathrm{m} (x_\mathrm{m}-x_\mathrm{b}) -\Gamma_\mathrm{m}(\dot{x}_\mathrm{m}-\dot{x}_\mathrm{b})\\
m_\mathrm{e} \ddot{x}_\mathrm{e} &=& -k_\mathrm{e} (x_\mathrm{e}-x_\mathrm{b}) -\Gamma_\mathrm{e}(\dot{x}_\mathrm{e}-\dot{x}_\mathrm{b})\\
m_\mathrm{b} \ddot{x}_\mathrm{b} &=& -k_\mathrm{c} (x_\mathrm{b}-x_\mathrm{0}) +k_\mathrm{m} (x_\mathrm{m}-x_\mathrm{b}) +k_\mathrm{e} (x_\mathrm{e}-x_\mathrm{b}) \nonumber\\
&& -\Gamma_\mathrm{c}(\dot{x}_\mathrm{b}-\dot{x}_\mathrm{0}) +\Gamma_\mathrm{m}(\dot{x}_\mathrm{m}-\dot{x}_\mathrm{b}) +\Gamma_\mathrm{e}(\dot{x}_\mathrm{e}-\dot{x}_\mathrm{b}).
\end{eqnarray}
Fig.~\ref{fig:model} shows the definitions of the employed parameters. 
$m_\mathrm{m}$, $m_\mathrm{e}$, and $m_\mathrm{b}$ are the test mass as well as the masses of the enclosure and base of the accelerometer, respectively. 
$k_\mathrm{m}$ and $\Gamma_\mathrm{m}$, $k_\mathrm{e}$ and $\Gamma_\mathrm{e}$, and $k_\mathrm{c}$ and $\Gamma_\mathrm{c}$ are the spring constants and damping coefficients between the test mass and base, enclosure and base, and base and measurement target, respectively. 
$x_\mathrm{m}$, $x_\mathrm{e}$, $x_\mathrm{b}$, and $x_\mathrm{0}$ are the inertial displacements of the test mass, enclosure, base, and measurement target, respectively. 
Notably, $k_\mathrm{c}$ and $\Gamma_\mathrm{c}$ include the contact stiffness and elastic deformation between the contact and reference points of $x_\mathrm{0}$.

Employing Fourier transformation, the equations are transformed into the frequency domain, as follows:
\begin{equation}
\left( \begin{array}{ccc} 
m_\mathrm{m} \kappa_\mathrm{m}(\omega)	&	0	&	-m_\mathrm{m}(\omega^2_\mathrm{m}+i\gamma_\mathrm{m}\omega)	\\
0	&	m_\mathrm{e}\kappa_\mathrm{e}(\omega)	&	-m_\mathrm{e}(\omega^2_\mathrm{e}+i\gamma_\mathrm{e}\omega)		\\
-m_\mathrm{m}\omega^2	&	-m_\mathrm{e}\omega^2	&	m_\mathrm{b}\kappa_\mathrm{c}(\omega)	\\
\end{array}\right) 
\left( \begin{array} {c}
\tilde{x}_\mathrm{m}\\
\tilde{x}_\mathrm{e}\\
\tilde{x}_\mathrm{b}
\end{array}\right) 
=
\left( \begin{array} {c}
0	\\
0	\\
m_\mathrm{b}(\omega^2_\mathrm{c}+i\gamma_\mathrm{c}\omega)
\end{array}\right) 
\tilde{x}_\mathrm{0},	\label{eq:eom_fd}
\end{equation}
where
\begin{eqnarray}
\kappa_\mathrm{m}(\omega) &=& \omega^2_\mathrm{m}+i\gamma_\mathrm{m}\omega-\omega^2	\\
\kappa_\mathrm{e}(\omega) &=& \omega^2_\mathrm{e}+i\gamma_\mathrm{e}\omega-\omega^2		\\
\kappa_\mathrm{c}(\omega) &=& \omega^2_\mathrm{c}+i\gamma_\mathrm{c}\omega-\omega^2.
\end{eqnarray}
Here, the angular resonance frequencies of the single spring, $\omega_\mathrm{m}\equiv\sqrt{k_\mathrm{m}/m_\mathrm{m}}$, $\omega_\mathrm{e}\equiv\sqrt{k_\mathrm{e}/m_\mathrm{e}}$, and $\omega_\mathrm{c}\equiv\sqrt{k_\mathrm{c}/m_\mathrm{b}}$ were introduced, and the damping constants were converted to $\gamma_\mathrm{m}=\Gamma_\mathrm{m}/m_\mathrm{m}$, $\gamma_\mathrm{e}=\Gamma_\mathrm{e}/m_\mathrm{e}$, and $\gamma_\mathrm{c}=\Gamma_\mathrm{c}/m_\mathrm{b}$.
In the above equations, $\tilde{x}$ represents the Fourier transform of $x$, and $\omega=2\pi f$ is the angular frequency.

The accelerometer outputs the electrical signal (charge, voltage, etc.) that is proportional to the relative displacement, $x_\mathrm{m}-x_\mathrm{b}$. 
In the conventional primary calibration, a reference laser interferometer is pointed to the upper surface of the target, as later shown in Fig.~\ref{fig:adaptercal_schematic}, to measure $x_0$.
Thereafter, the sensitivity is calculated as the ratio of the accelerometer output to the reference acceleration, $\ddot{x}_\mathrm{0}$; and it is expressed, as follows:
\begin{equation}
S(\omega) = G_\mathrm{el} \frac{\tilde{x}_\mathrm{m}-\tilde{x}_\mathrm{b}}{-\omega^2\tilde{x}_\mathrm{0}},
\label{eq:S_definition}
\end{equation}
where $G_\mathrm{el}$ is the conversion gain from the relative displacement to the electrical signal.
By solving Eq.~(\ref{eq:eom_fd}),
\begin{equation}
S(\omega) = S_\mathrm{0} \frac{ (\omega^2_\mathrm{c}+i\gamma_\mathrm{c}\omega) \omega^2_\mathrm{m} \kappa_\mathrm{e} }{ \kappa_\mathrm{m} \kappa_\mathrm{e} \kappa_\mathrm{c} - (\omega^2_\mathrm{m}+i\gamma_\mathrm{m}\omega) \omega^2 \eta^{-1}_\mathrm{c} \kappa_\mathrm{e} - (\omega^2_\mathrm{e}+i\gamma_\mathrm{e}\omega) \omega^2 \eta_\mathrm{e} \eta^{-1}_\mathrm{c} \kappa_\mathrm{m} } \label{eq:S}
\end{equation}
becomes the sensitivity obtained from the conventional primary calibration for SE accelerometers.
Here, the mass ratios were defined as $\eta_\mathrm{c}\equiv m_\mathrm{b}/m_\mathrm{m} $ and $\eta_\mathrm{e}\equiv m_\mathrm{e}/m_\mathrm{m} $.
In this study, the low-frequency sensitivity is defined as $S_\mathrm{0}\equiv G_\mathrm{el} \omega^{-2}_\mathrm{m}$.
Although $S_\mathrm{0}$ typically exhibits high-pass characteristics with a cutoff of approximately 1~Hz attributed to the charge-amplifier settings, it is considered a constant in this study because of our focus onhigh frequencies above 100~Hz. 

Assuming that the contact between the accelerometer and target is completely rigid, i.e., in the limit of $k_\mathrm{c}\to\infty$, the sensitivity would become
\begin{equation}
S(\omega) \xrightarrow[k_\mathrm{c}\to\infty]{} S_\mathrm{0} \frac{\omega^2_\mathrm{m}}{\omega^2_\mathrm{m}+i\gamma_\mathrm{m}\omega-\omega^2}.
\label{eq:S_rigid}
\end{equation}
In this ideal case, the sensitivity depends only on the test mass parameters ($\omega_\mathrm{m}$ and $\gamma_\mathrm{m}$) and exhibits a resonance peak at $\omega_\mathrm{m}$.
However, in the real system, additional resonant peaks appear because the base, enclosure, and test mass form the coupled spring.
As expressed in Eq.~(\ref{eq:S}), the real frequency response depends on the mounting condition ($k_\mathrm{c}$ and $\Gamma_\mathrm{c}$).

\section{Calibration methods} \label{sec:method}
\subsection{Reversed calibration for the measurement of the intrinsic sensitivity} \label{sec:reversed}
\begin{figure}
	\begin{center}
	\includegraphics[width=10cm]{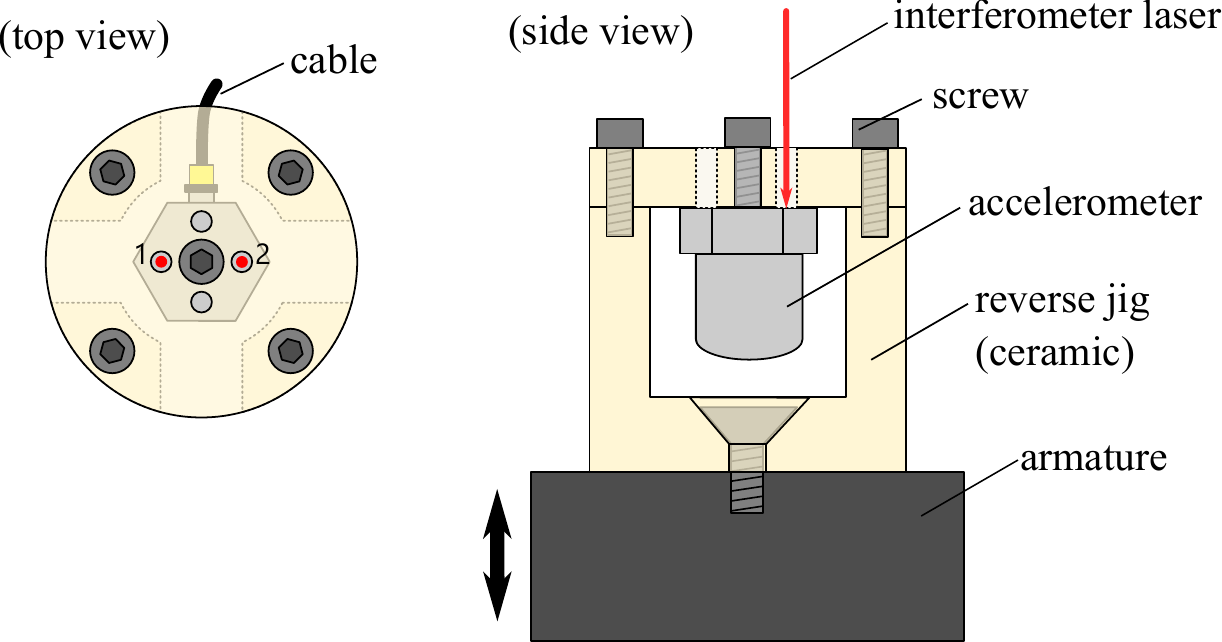}
	\caption{Schematic view of the reversed calibration. A SE accelerometer is fixed upside-down, and the laser interferometer beam is directly pointed at the accelerometer surface through the holes of the jig plate. Among the four holes, two holes labeled as position 1 and 2 were used for the measurement.}
	\label{fig:revcal_schematic}
	\end{center}
\end{figure}
Fig.~\ref{fig:revcal_schematic} shows the schematic view of the reversed calibration proposed in this study.
The accelerometer under test was set upside down on the vibration exciter using the reverse jig, which is made of ceramic (alumina) to achieve high stiffness with light weight.
The reverse jig weighed 215~g, which is within the load capacity of the vibration exciter, SE-09.
Dissimilar to the conventional primary calibration, the laser beam of the reference interferometer was pointed to the surface of the accelerometer base through the hole in the jig plate.
In this case, the sensitivity was measured with respect to the base acceleration, $\ddot{x}_\mathrm{b}$.
Therefore, based on the calculation in Section~\label{sec:model}:
\begin{equation}
S_\mathrm{rev}(\omega) = G_\mathrm{el} \frac{\tilde{x}_\mathrm{m}-\tilde{x}_\mathrm{b}}{-\omega^2\tilde{x}_\mathrm{b}} = S_\mathrm{0} \frac{\omega^2_\mathrm{m}}{\omega^2_\mathrm{m}+i\gamma_\mathrm{m}\omega-\omega^2}.
\label{eq:S_intrinsic}
\end{equation}
We considered $S_\mathrm{rev}$ an ``intrinsic sensitivity'' throughout this paper, as it is ideally independent of the mounting condition.
This is identical to the sensitivity in the case of completely rigid mounting in the adapter calibration (Eq.~(\ref{eq:S_rigid})).

We employed a standard SE accelerometer B\&K 8305-001 for this experiment. 
The calibration method followed ISO 16063-11. 
Sinusoidal vibration was applied to the accelerometer using SE-09 (Spektra), and its output signal was compared with that of a reference heterodyne laser interferometer LV-9002 (OnoSokki) to calculate the sensitivity \cite{Kokuyama2022}.
The measurement frequency band was between 100~Hz and 95~kHz, with 500~Hz increments above 4~kHz.
Although the specified frequency range of SE-09 was up to 50~kHz, we drove it above 50~kHz using moderate-acceleration amplitude. 
The charge signal of the accelerometer was converted into voltage using a charge amplifier (B\&K 2692), after which it was measured by a digitizer PXI-5922 (National Instruments). 
The low-pass cutoff of the charge amplifier was set to 100 kHz.
The charge-amplifier gain (in V/pC) was calibrated independently, and the recorded voltage was converted into a charge signal.

\begin{figure}
	\begin{center}
	\includegraphics[width=10cm]{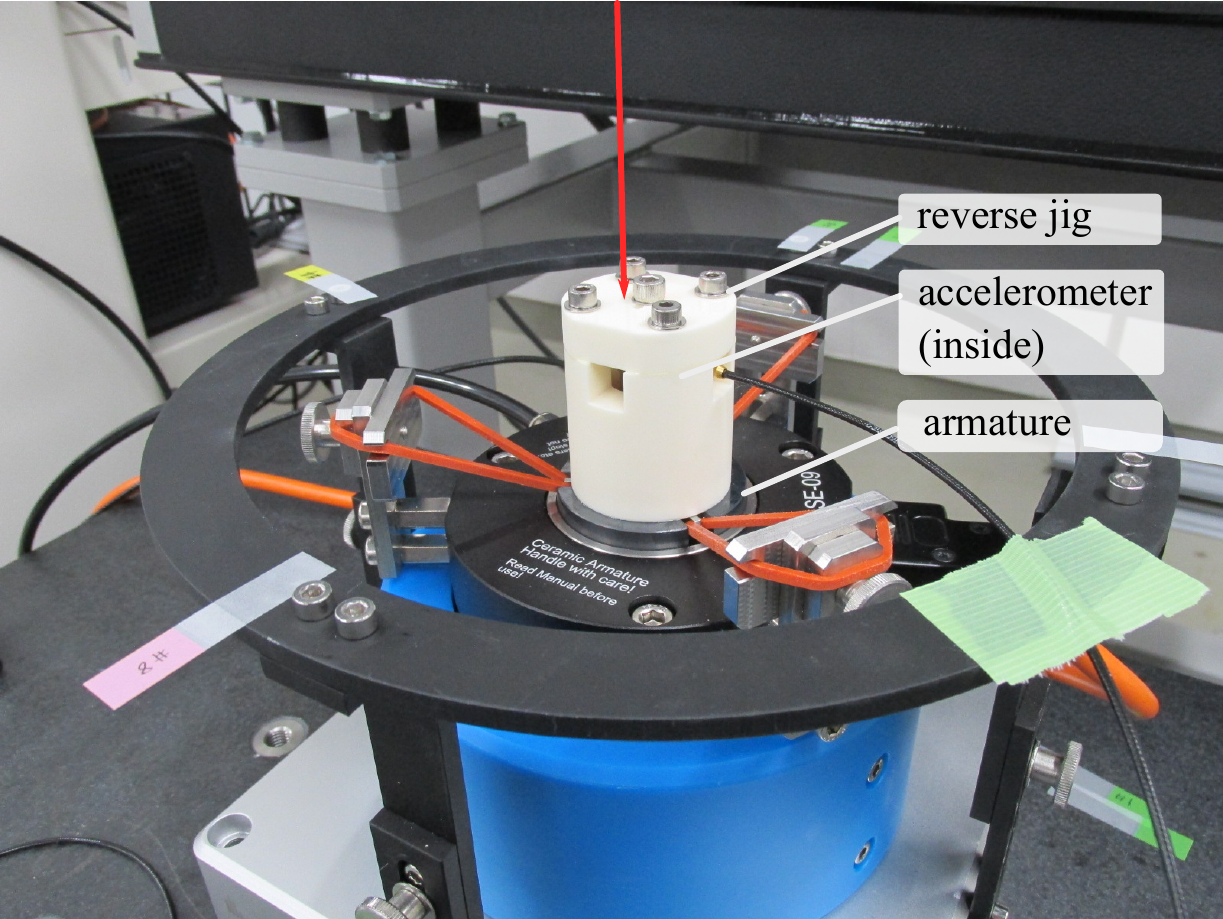}
	\caption{Experimental setup for reversed calibration. The reverse jig is a milky-white cylinder.}
	\label{fig:revcal_setup}
	\end{center}
\end{figure}

The measurement results obtained at Positions 1 and 2 (Fig.~\ref{fig:revcal_schematic}) were averaged to cancel the armature-tilting effect. 
The hole closest to the cable connector was not utilized to avoid the effect of the local deformation of the accelerometer body due to the connector mass. 
The jig plate was fixed to the jig base using four screws, and the base is fixed to the armature of the vibration exciter with a screw at the center. 
Here, grease was applied between the accelerometer and the jig, and the mounting torque was set to 2~N$\cdot$m.
Fig.~\ref{fig:revcal_setup} shows the actual setup.

\subsection{Calibration with adapters} \label{sec:adapter}
\begin{figure}
	\begin{center}
	\includegraphics[width=10cm]{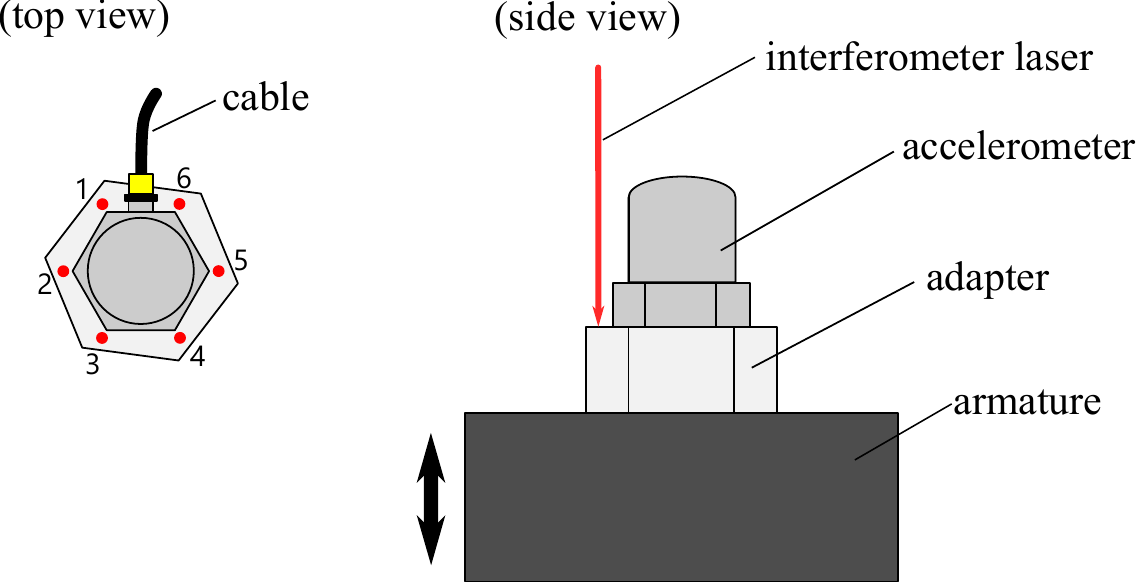}
	\caption{Schematic view of the conventional calibration using adapter. An accelerometer is fixed to the hexagonal prism adapter, and the laser interferometer beam is pointed at the adapter surface. The laser was pointed at six points close to the hexagonal vertices of the accelerometer.}
	\label{fig:adaptercal_schematic}
	\end{center}
\end{figure}
We also performed the conventional adapter calibration with four different materials up to 100~kHz for two reasons. 
The intrinsic-sensitivity parameters can be estimated by analyzing the adapter-calibration results based on the three-body model; hence, the result independently verifies the performed reversed calibration. 
Additionally, the quantitative information regarding material dependency offers valuable insights for discussing high-frequency calibration and measurement uncertainties.

Fig.~\ref{fig:adaptercal_schematic} shows the schematic of the calibration using the adapter. 
The configuration is the same as that adopted in the international comparison, CCAUV.V-K5. 
The employed accelerometer was attached to the adapter using a set screw, and the adapter was set to the armature, and the laser beam of the reference interferometer was pointed to the upper surface of the adapter.
The measured sensitivity is expressed by Eq.~(\ref{eq:S}), where the contact parameters ($k_\mathrm{c}$ and $\Gamma_\mathrm{c}$) were determined by combining the accelerometer and adapter. 
As the denominator and numerator of Eq.~(\ref{eq:S}) are sixth- and second-order polynomials of $\omega$, respectively, the frequency response was expected to exhibit three resonance peaks and one anti-resonant valley.

The same standard SE accelerometer B\&K 8305-001 as used for the reversed calibration was deployed. 
The signal-acquisition setup, as well as the processing, were also similar to the reversed calibration explained in Section~\ref{sec:reversed}. 
Six measurement results obtained at Positions 1--6 (Fig.~\ref{fig:adaptercal_schematic}) were averaged to cancel the armature-tilting effect. 
To improve the contact, grease was applied between the accelerometer and adapter, and the mounting torque was set to 2~N$\cdot$m. 
The upper surface of each adapter was polished to a surface roughness of $R_\mathrm{a}<2$~nm. 
In this experiment, we employed four adapters, which were made of tungsten carbide, stainless steel, titanium, and aluminum, respectively. 
The typical Young's moduli of these materials are 600, 193, 106, and 68~GPa, respectively.

\section{Measurement result}	\label{sec:result}
\subsection{Intrinsic sensitivity measured by reversed calibration} \label{sec:result.reverse}
\begin{figure}
	\begin{center}
	\includegraphics[width=10cm]{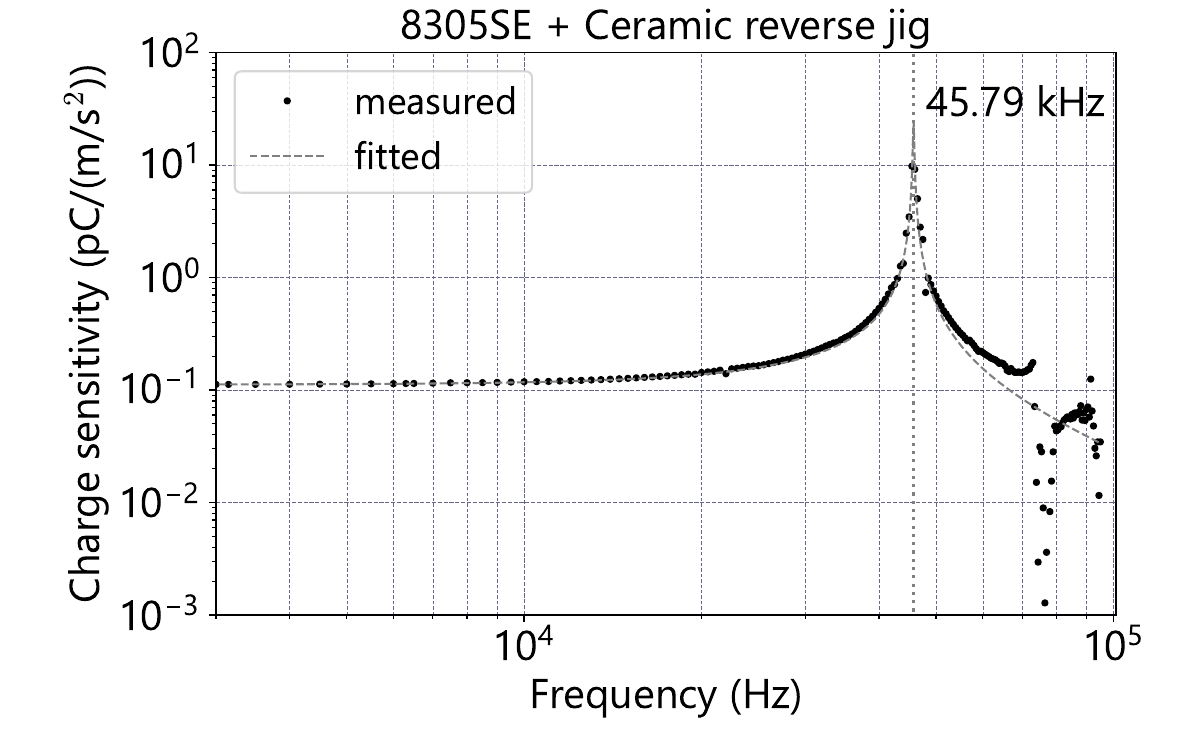}
	\caption{Measurement result of the reversed calibration. The black circles and the dashed grey line are the measured sensitivity and the fitting result, respectively.}
	\label{fig:result_revcal}
	\end{center}
\end{figure}
Fig.~\ref{fig:result_revcal} shows the reversed-calibration results. 
As anticipated from the model, the sensitivity magnitude displayed one main resonance peak at 45.8~kHz, corresponding to the intrinsic resonance frequency, $\omega_\mathrm{m}/(2\pi)$. 
The unexpected frequency structure around 75~kHz might be due to the coupling from the enclosure deformation, as its resonance frequency was around 75~kHz (see Section~\ref{sec:result.adapter}). 
Table~\ref{table:params0} lists the parameters that were estimated by fitting the data using Eq.~(\ref{eq:S_intrinsic}). 
In the fitting, the low-frequency sensitivity, $S_0$, was fixed to the average of the measured value between 100~Hz and 300~Hz; additionally, $\omega_\mathrm{m}$ and $\gamma_\mathrm{m}$ were treated as fitting parameters.
Each data point was equally weighted on a logarithmic scale for simplicity, i.e., the relative errors were assumed to be approximately uniform. 
Furthermore, the measured peak frequency, 45.8~kHz, was employed as the initial estimate of $\omega_\mathrm{m}/(2\pi)$.
\begin{table}
\begin{center}
\caption{Estimated parameters from the reversed calibration}\label{table:params0}
\begin{tabular}{cc|c}
	Parameter	& Unit	& Value	\\ \hline 
	$S_0$	& pC/(m/s$^2$)	& 0.11144 \\ 
	$\omega_\mathrm{m}$	& $\mathrm{s}^{-1}$	& $2\pi\times45789$ \\ 
	$\gamma_\mathrm{m}$	& $\mathrm{s}^{-1}$	& 1311 \\ 
\end{tabular}
\end{center}
\end{table}

\subsection{Calibrated sensitivity with adapters} \label{sec:result.adapter}
\begin{figure}
\begin{minipage}[t]{1\hsize}
	\begin{center}
	(a)\\
	\includegraphics[width=10cm]{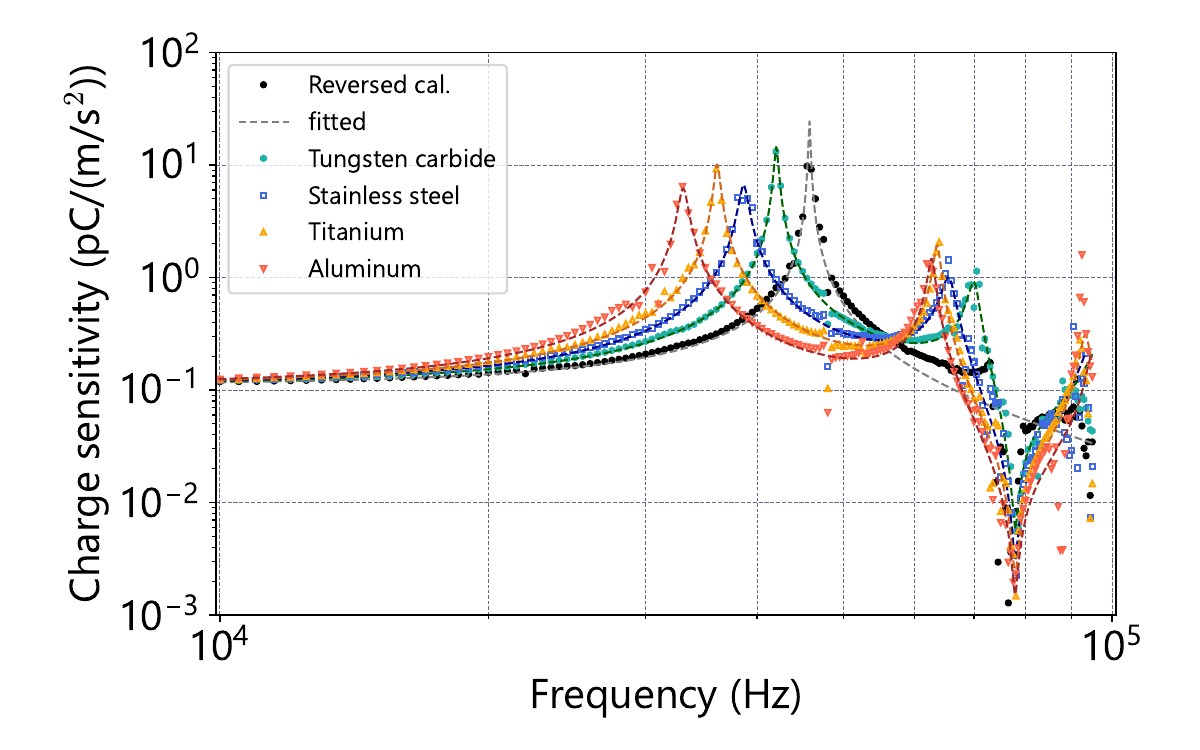}
	\end{center}
\end{minipage}
\begin{minipage}[t]{1\hsize}
	\begin{center}
	(b)\\
	\includegraphics[width=10cm]{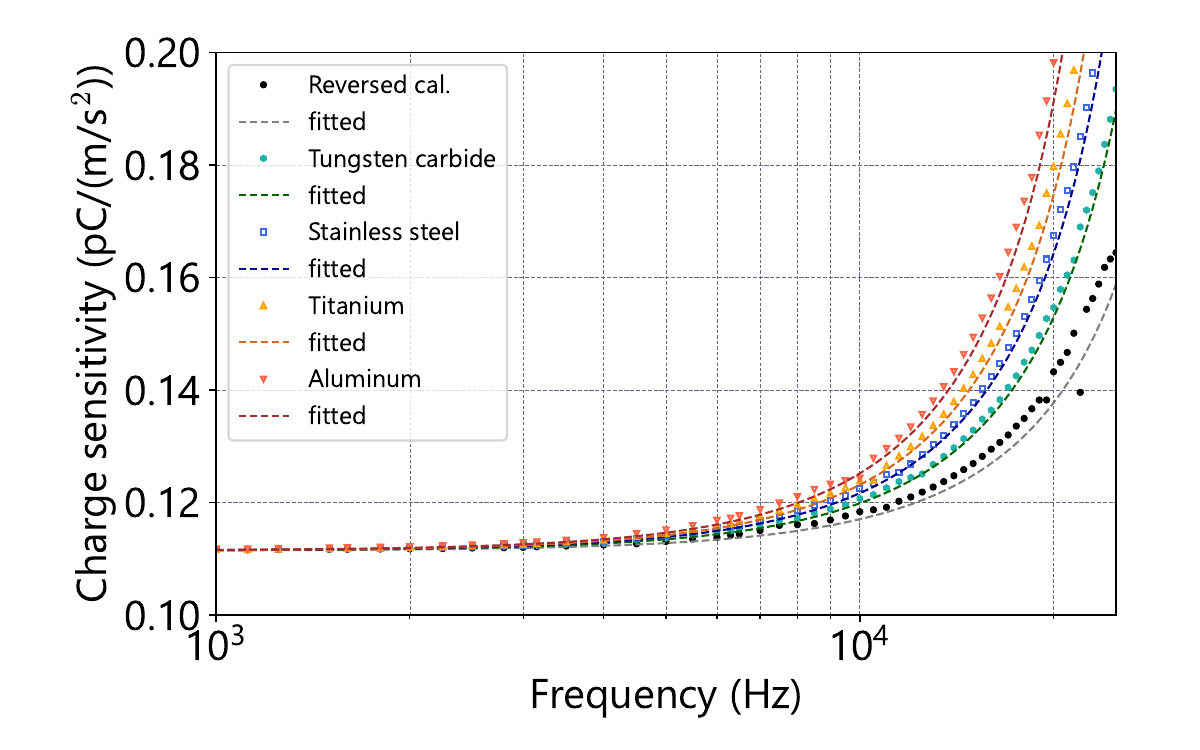}
	\end{center}
\end{minipage}
\caption{Measured sensitivities with adapters (a) above 10~kHz and (b) below 25~kHz. The reversed calibration result is plotted with black circle, and the adapter calibration results are plotted with green hexagon (tungsten carbide), light blue cross (tungsten), blue square (stainless steel), orange triangle (titanium), and red triangle (aluminum). Fitting results are also plotted with dashed lines.}
\label{fig:result_adapter}
\end{figure}
Fig.~\ref{fig:result_adapter} shows the calibration results using the adapters. 
As anticipated from Eq.~(\ref{eq:S}), the frequency responses displayed three resonance peaks and one anti-resonance valley at 78~kHz. 
However, for some adapters, it was challenging to observe the highest resonance peak around 90~kHz owing to the measurement noise. 
The lowest resonant frequencies ranged between 33~kHz and 46~kHz depending on the stiffness of the adapter.
Eq.~(\ref{eq:S}) indicates that the anti-resonance at 78~kHz corresponded to the resonance of the enclosure, $\omega_\mathrm{e}/2\pi$, which minimized the factor, $\kappa_\mathrm{e}$, in the numerator.

The parameters were estimated by fitting the data using Eq.~(\ref{eq:S}). 
$S_0$ was fixed to the same value as employed in the reversed calibration, and the other parameters were treated as fitting parameters. 
Here, the four sets of sensitivity data obtained with different adapters were fitted concurrently. 
The five parameters, $\omega_\mathrm{m}$, $\gamma_\mathrm{m}$, $\eta_\mathrm{e}$, $\omega_\mathrm{e}$, and $\gamma_\mathrm{e}$, were set common for all the adapters, whereas $\eta_\mathrm{c}$, $\omega_\mathrm{c}$, and $\gamma_\mathrm{c}$ were assigned different values for each adapter. 
Put differently, 17 fitting parameters (five common parameters and three individual parameters for four adapters) were estimated in a single fitting process, and the estimation results are presented in Table~\ref{table:params}. 
The model sensitivities obtained with the fitted parameters are shown in Fig.~\ref{fig:result_adapter} using dashed lines, which well explain the frequency dependence of the measured sensitivities.

As the damping coefficients mainly affected the sharpness of the resonance peaks, the overall change in the frequency response might have been caused by the differences between $\omega_\mathrm{c}$ and $\eta_\mathrm{c}$, or $k_\mathrm{c}$ and $m_\mathrm{b}$. 
The difference in $k_\mathrm{c}$ represents the change in the contact stiffness and elastic deformation. Notably, $\eta_\mathrm{c}$ is ideally independent of the adapter materials, although its material dependency was considered in this analysis, as the effective mass of the elastic deformation mode might exhibit different values for different materials. 
Indeed, accounting for this material dependency improved the fitting results, particularly around 60 kHz.

\begin{table}
\begin{center}
\caption{Estimated parameters from the calibration with different adapters}\label{table:params}
\begin{tabular}{cc|cccc}
	Parameter	& Unit	& Tungsten carbide	& Stainless steel	& Titanium	& Aluminum	\\ \hline 
	$\omega_\mathrm{m}$	& $\mathrm{s}^{-1}$	& \multicolumn{4}{c}{$2\pi\times45912$} \\ 
	$\gamma_\mathrm{m}$	& $\mathrm{s}^{-1}$	& \multicolumn{4}{c}{400} \\ 
	$\eta_\mathrm{e}$	& \textendash	& \multicolumn{4}{c}{0.295} \\ 
	$\omega_\mathrm{e}$	& $\mathrm{s}^{-1}$	& \multicolumn{4}{c}{$2\pi\times77845$} \\ 
	$\gamma_\mathrm{e}$	& $\mathrm{s}^{-1}$	& \multicolumn{4}{c}{5393} \\ 
	$\eta_\mathrm{c}$	& \textendash	& 2.460	& 1.919	& 1.696	& 1.361	 \\ 
	$\omega_\mathrm{c}$	& $\mathrm{s}^{-1}$	& $2\pi\times80487$	& $2\pi\times66646$	& $2\pi\times59985$	& $2\pi\times55201$	 \\ 
	$\gamma_\mathrm{c}$	& $\mathrm{s}^{-1}$	& 41686	& 40193	& 17377	& 21304	 \\ 
\end{tabular}
\end{center}
\end{table}

\begin{figure}
	\begin{center}
	\includegraphics[width=10cm]{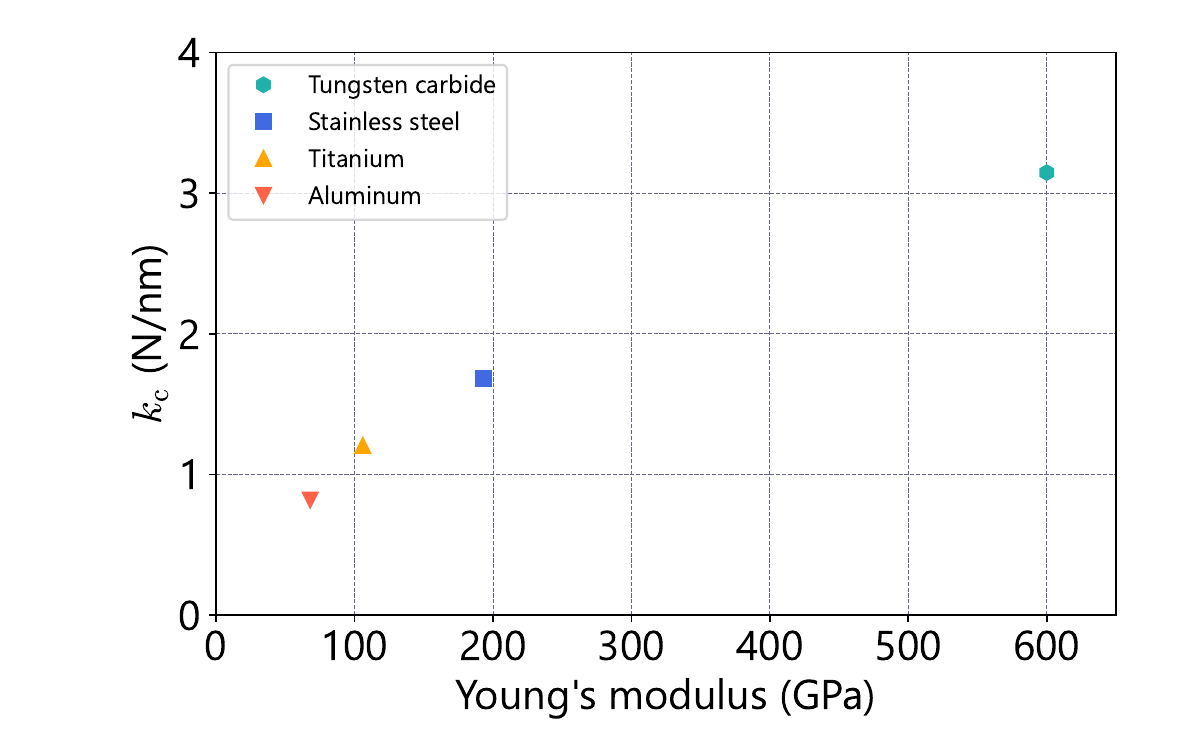}
	\end{center}
\caption{Dependence of the parameter $k_\mathrm{c}$ calculated from the fitted $\omega_\mathrm{c}$ and $\eta_\mathrm{c}$. The mass $m_\mathrm{m}$ was roughly assumed to be 5~g. The parameters are plotted with green hexagon (tungsten carbide), blue square (stainless steel), orange triangle (titanium), and red triangle (aluminum).}
\label{fig:result_params}
\end{figure}
Fig.~\ref{fig:result_params} shows the change in $k_\mathrm{c}$ calculated from Table~\ref{table:params} depending on the Young's modulus of the adapter. 
Notably, $k_\mathrm{c}$ is a series spring of the contact stiffness and elastic deformation. 
The spring constant of the elastic deformation was expected to be roughly proportional to the Young modulus, whereas the dependency of the contact stiffness was complicated and unclear. 
Therefore, the nonlinear dependence illustrated in Fig.~\ref{fig:result_params} indicates that the contact stiffness and elastic deformation contributed to the sensitivity.

\section{Discussion and conclusion}	\label{sec:discussion}
The reversed-calibration method for measuring the intrinsic sensitivity of an accelerometer was experimentally demonstrated in this study. 
The calibrated sensitivity exhibited a single main resonance peak, as anticipated from the model, and the identified intrinsic resonance frequency was 45.79~kHz (Fig.~\ref{fig:result_revcal}). 
For the independent verification, the same parameter was estimated by fitting the adapter-calibration results (Table~\ref{table:params}), exhibiting a similar value as $\omega_\mathrm{m}/2\pi=45.91$~kHz. 
The correlation between the two independent measurements supported the correctness of reversed calibration as well as the validity of the three-body model. 
Although the detailed calibration uncertainties are still under investigation, the fundamental measurement concept has been validated in this study.

In detail, the measured sensitivity differed from the model by up to 4~\% at 20~kHz (Fig.~\ref{fig:result_adapter}~(b)) owing to the contributions that were not considered in the model, such as the higher-order modal deformation of the components, signal-output coupling with deformation other than the seismic system, and transverse vibration. 
For example, the deviation between the measured and expected intrinsic sensitivities (Fig.~\ref{fig:result_revcal}) at the enclosure resonance around 78~kHz indicated the existence of coupling between the enclosure deformation and accelerometer output. 
Such a coupling can exhibit a low-frequency tail that affects the sensitivity around 20~kHz.

Notably, we observed a change in the resonance structure by performing adapter calibration above 50~kHz for the first time. 
The measured frequency response and its change with the adapter materials were well explained by the three-body model (Fig.~\ref{fig:result_params}~(a)). 
By the fitting performed in Section~\ref{sec:result.adapter}, we confirmed that the parameters related to the contact part, $k_\mathrm{c}$ and $m_\mathrm{b}$, were key to the sensitivity changes. 
A similar conclusion was reported by Ref. \cite{Bruns2012} who investigated the effect of $k_\mathrm{c}$ based on the two-body model. 
Our results support the fundamental understanding of the extant study and reveal the detailed mechanism by improving the dynamic model as well as measuring the sensitivity above 50~kHz. 
The sensitivity difference between the stainless-steel and aluminum adapters, which are often employed in industries, was 20~\% at 20~kHz (Fig.~\ref{fig:result_adapter}), indicating that when an accelerometer calibrated with a stainless-steel adapter is attached to the aluminum body, the measured vibration can exhibit an amplitude error of 20~\% around 20~kHz.

The reversed-calibration method could be applied to other accelerometers. 
As the next step of our research, the evaluation of the behaviors of different accelerometers would be key to elucidating the general degree of the effect of the mounting conditions. 
As the accelerometers deployed in actual applications are typically calibrated through a series of primary and comparison calibrations, the uncertainty due to the mounting effect at each step is transferred to the final accelerometer sensitivity. 
Notably, discussions on the degree of calibration uncertainty and its suppression technique will be crucial in vibration metrology. 
The methodology reported in this paper will serve as a valuable approach for obtaining the necessary information for such discussions.

\section*{Acknowledgements}
The authors thank Hiromi Mitsumori (NMIJ) for technical support in the experiment.

\section*{Author contributions: CRediT}
\textbf{Tomofumi Shimoda:} Conceptualization, Methodology, Investigation, Formal analysis, Resources, Software, Visualization, Funding acquisition, Writing – original draft
\textbf{Wataru Kokuyama:} Methodology, Resources, Software, Supervision, Writing – review and editing
\textbf{Hideaki Nozato:} Methodology, Supervision, Funding acquisition, Writing – review and editing

\section*{Funding sources}
This work was supported by JSPS KAKENHI Grant Number JP22K14273.


\begin{thebibliography}{00}
\bibitem{ISO16063-11} International Organization for Standardization
    \newblock{ISO 16063-11:1999, Methods for the calibration of vibration and shock transducers - Part 11: Primary vibration calibration by laser interferometry}
\bibitem{Dobosz1997I} M. Dobosz, T. Usuda, and T. Kurosawa
    \newblock{Methods for the calibration of vibration pick-ups by laser interferometry: I. Theoretical analysis}, {\it Measurement Science and Technology}, {\bf 9}, 2, 232 (1998). \url{https://doi.org/10.1088/0957-0233/9/2/010}
\bibitem{Dobosz1997II} M. Dobosz, T. Usuda, and T. Kurosawa
    \newblock{Methods for the calibration of vibration pick-ups by laser interferometry: II. Experimental verification}, {\it Measurement Science and Technology}, {\bf 9}, 2, 240 (1998). \url{https://doi.org/10.1088/0957-0233/9/2/011}
\bibitem{Martens2000} H. J. von Martens, A. T\"{a}ubner, W. Wabinski, A. Link, and H. J. Schlaak
    \newblock{Traceability of vibration and shock measurements by laser interferometry}, {\it Measurement}, {\bf 28}, 1, 3 (2000). \url{https://doi.org/10.1016/s0263-2241(00)00003-8}
\bibitem{Martens2013} H. J. von Martens
    \newblock{Invited Article: Expanded and improved traceability of vibration measurements by laser interferometry}, {\it Review of Scientific Instruments}, {\bf 84}, 121601 (2013). \url{https://doi.org/10.1063/1.4845916}
\bibitem{Winther2022} J. H. Winther, M. Brucke, M. T. Andersen, and T. R. Licht
    \newblock{Recent study in primary accelerometer calibration -- progress, development and lessons learned}, {\it proceedings of IMEKO 5th International Conference} (2022). \url{https://doi.org/10.21014/tc22-2022.081}
\bibitem{Kokuyama2022} W. Kokuyama, T. Shimoda, and H. Nozato
	 \newblock{Primary accelerometer calibration with two-axis automatic positioning stage}, {\it Measurement}, {\bf 204}, 30, 112044 (2022). \url{https://doi.org/10.1016/j.measurement.2022.112044}
\bibitem{CCAUV.V-K2} T. Bruns, G. P. Ripper, and A. T\"{a}ubner
    \newblock{Final report on CIPM key comparison CCAUV.V-K2}, {\it Metrologia}, {\bf 51}, 1A, 09002 (2014). \url{https://doi.org/10.1088/0026-1394/51/1a/09002}
\bibitem{CCAUV.V-K5} T. Bruns {\it et al.} 
    \newblock{Final report on the CIPM key comparison CCAUV.V-K5}, {\it Metrologia}, {\bf 58}, 1A, 09001 (2021)
\bibitem{Ripper2013} G. P. Ripper, G. B. Micheli, and R. S. Dias
    \newblock{A proposal to minimize the dispersion on primary calibration results of single-ended accelerometers at high frequencies}, {\it Acta IMEKO}, {\bf 2}, 2, 48 (2013). \url{https://doi.org/10.21014/acta_imeko.v2i2.84}
\bibitem{Taubner2010} A. T\"{a}ubner, H. Schlaak, M. Brucke, and T. Bruns
    \newblock{The influence of different vibration exciter systems on high frequency primary calibration of single-ended accelerometers}, {\it Metrologia}, {\bf 47}, 1, 58 (2010). \url{https://doi.org/10.1088/0026-1394/47/1/007}
\bibitem{Bruns2012} T. Bruns, A. Link, and A. T\"{a}ubner
    \newblock{The influence of different vibration exciter systems on high frequency primary calibration of single-ended accelerometers: II}, {\it Metrologia}, {\bf 49}, 1, 27 (2012). \url{https://doi.org/10.1088/0026-1394/49/1/005}
%\bibitem{Andreas2022} A. Havreland
%	\newblock{Electrical impedance based calibration of accelerometers}, {\it Proceedings of IMEKO 5th TC22 International Conference} (2022)
\end{thebibliography}
\end{document}